\documentclass{article}
\usepackage{graphicx}
\usepackage{amsmath}

\begin{document}

\begin{titlepage}

\begin{center}

\vspace{1cm}

{\Large \textbf{Gravitational Higgs Mechanism: The Role of
Determinantal Invariants}}

\vspace{1cm}

{\bf Nurettin Pirin{\c c}{\c c}io{\~g}lu\\}

\vspace{0.5cm}

{\it Department of Physics, University of Dicle, Diyarbak\i r,
TR21280.\\
(August, 2009)}

\end{center}

\vspace{1cm}

\begin{abstract}

The Higgs mechanism for gravity, as proposed by 't Hooft in arXiv:
0708.3184[hep-th], can be augmented by including determinantal
invariants. We analyze the effects of determinantal invariants in
such a set up. We find that the part of the potential that depends
on the determinantal invariants, if obtains a specific exponential
form in terms of its argument, may not affect the graviton mass
calculated.

\end{abstract}

\vspace{8cm}

\begin{flushleft}
{Department of Physics,\\
Faculty of Science,\\
University of Dicle,\\
Diyarbak{\i }r, 21280 TURKEY\\
npirinc@dicle.edu.tr}
\end{flushleft}

\end{titlepage}

\section{Introduction}

In gauge theories, Higgs mechanism is the most fundamental
framework for generating gauge boson masses in a renormalizable
and continuous fashion. Namely, in non-Abelian gauge theories the
discontinuity in physical amplitudes as gauge boson mass tends to
vanish is lifted by the Higgs mechanism \cite{vainshtein}. There
has been a recent activity as to derive a Higgs mechanism for
gravity such that discontinuity in graviton mass \cite{iwasaki} is
lifted by the associated Higgs mechanism. To this end, Moffat
proposed the spontaneous symmetry breaking of local Lorentz
invariance \cite{moffat}. 't Hooft \cite{hooft2007} studied the
Brout-Englert Higgs mechanism using four scalar field to fix the
gauge in the four-dimensional spacetime. Scalars play the role of
preferred frame. Later Kakushadze \cite{zurab2007} derived an
equation for Pauli-Fierz mass term in terms of a general potential
by using Einstein-Hilbert action to the linear order. Very
recently, Demir and Pak \cite{durmush2009} discovered an equation
for the mass term by using a more general action to quadratic
order. Their action, compared to the previous ones, contains also
determinantal invariants. There are also several other studies
considering gravitational Higgs mechanism in different contexts
\cite{zurabpeter}, \cite{poratti}, \cite{ali}, \cite{kirsch},
\cite{berezhiani2007}, \cite{nima2002}, \cite{rubakov2008},
\cite{chamseddine}.

The Pauli-Fierz theory \cite{fierz1939}, which is nothing but
linearized Einstein-Hilbert action about flat Minkowski, is the
only ghost (negative energy) free theory.  For this theory the
mass term in the lagrangian must have the form
\begin{equation}
-\frac{m^{2}_{g}}{4}(h_{\mu\nu}h^{\mu\nu}-\xi h^{2})
\end{equation}
where $h=h^{\mu}_{\mu}$ is the non-unitary, negative energy mode,
which is eliminated  for $\xi=1$.

The framework of the present paper is that of Demir and Pak
\cite{durmush2009}. Our goal in this paper is to analyze the
effects of determinantal potential on the graviton mass term. We
will show that, a physically admissible mass term arises for a
specific solution for the determinantal potential.

To get ready for the analysis in the next section, we want to lay
out notation used: We consider 4-dimensional spacetime with
coordinates $x^{\mu}$. In addition, we introduce scalar
coordinates $\phi^{a}$ $(a=0, 1, 2, 3)$ parallel to $x^{\mu}$. In
addition to the spacetime metric $g_{\mu\nu}£$, we introduce a
scalar-induced metric
\begin{equation}
g^{\phi}_{\mu\nu}=\eta_{ab}\partial_{\mu}\phi^{a}\partial_{\nu}\phi^{b}
\end{equation}
where $\eta_{ab}=diag(-1, 1, 1, 1)$ is the flat Minkowski metric,
not the identity matrix, i.e., $\eta_{ab}\neq\delta_{ab}$.

Incidentally, the determinant of $g^{\phi}_{\mu\nu}$ reads to be
\begin{equation}
\sqrt{\mbox{-det}(g^{\phi}_{\mu\nu})}=
\varepsilon^{\alpha\beta\gamma\rho}\partial_{\alpha}\phi^{0}\partial_{\beta}\phi^{1}\partial_{\gamma}\phi^{2}\partial_{\rho}\phi^{3}
\end{equation}
where $\varepsilon^{\alpha\beta\gamma\rho}$ is Levi-Civita symbol.
One here observes that
$\mbox{det}(g^{\phi}_{\mu\nu})/\mbox{det}(g_{\mu\nu})$ is a
perfect scalar field. This is the basis of the work
\cite{durmush2009}, and our aim in this work is to analyze the
potential of this scalar field, i.e. ratio of the determinants.

In weak gravitational field, one can expand $g_{\mu\nu}$ as
\begin{equation}
g_{\mu\nu}=\eta_{\mu\nu}+h_{\mu\nu}
\end{equation}
with inverse
\begin{equation}
g^{\mu\nu}=\eta^{\mu\nu}-h^{\mu\nu}+\frac{1}{2}h^{\mu\alpha}h_{\alpha}^{\nu}
\end{equation}
up to the quadratic order. One obtains
\begin{equation}
g=\mbox{det}(g_{\mu\nu})=-1-h.
\end{equation}
The Higgs mechanism of 't Hooft is based on the fact that the
scalar coordinates acquire the vacuum expectation value
\begin{equation}
\langle{\phi^{a}}\rangle=m^{2}x^{a}
\end{equation}
so that
\begin{equation}
\langle{g^{\phi}_{\mu\nu}}\rangle=m^{4}\eta_{\mu\nu}
\end{equation}
where $m$ is the mass scale. Therefore, in the vacuum, the
determinant of metric, $g$, is a scalar density of weight $+2$.
The flat induced metric, its inverse, and its determinant with the
same argument can be defined respectively as follows

\begin{equation}
\langle{g^{\phi}_{\mu\nu}}\rangle=m^{4}\eta_{\mu\nu},
\end{equation}
\begin{equation}
\langle{g^{(\phi)\mu\nu}}\rangle=\frac{1}{m^{4}}\eta^{\mu\nu},
\end{equation}
\begin{equation}
\langle{g^{\phi}}\rangle=\langle\mbox{det}(g^{\phi}_{\mu\nu})\rangle=-m^{16}.
\end{equation}

The outline of the paper is as follows. The action, which includes
the Einstein-Hilbert term in addition to the kinetic potential
\cite{zurab2007} and the potential of the determinantal invariant
is introduced in Section $2$. From this action, constraints on the
potentials in the massless phase is given in Section $3$, and the
linearized equation of motion for massive graviton derived in
Section $4$. Finally, it is completed with a conclusion in Section
$5$.

\section{Action for graviton}

From the set up of  \cite{zurab2007} and \cite{durmush2009} one
can consider the following action:
\begin{equation}
S=M_{p}^{D-2}\int d^{D}x\sqrt{-g}\left[ R-V(y)+F(f)\right]   \label{Action}
\end{equation}
where $R=g_{\mu\nu}R^{\mu\nu}$ is the curvature scalar, and
\begin{equation}
y=g^{\phi}_{\mu\nu}g^{\mu\nu},
\end{equation}
\begin{equation}
f=\mbox{det}(g^{\phi}_{\mu\nu})/\mbox{det}(g_{\mu\nu})
\end{equation}
are kinetic and determinantal invariants, respectively. 't Hooft's
work considers only the invariant $y$. Demir and Pak include the
invariant $f$, too. In this paper all discussions will be in the
four dimensional spacetime, $D=4$. Thus, the variation of equation
(\ref{Action}) with respect to the metric $g_{\mu\nu}$ gives

\begin{equation*}
(R_{\mu\nu}-\frac{1}{2}g_{\mu\nu}R)=
\end{equation*}
\begin{equation}
 V'(y)g^{\phi}_{\mu\nu}+\frac{1}{2 }g_{\mu\nu}\left[
F(f)+2fF'(f)-V(y)\right]   \label{metricmotion}
\end{equation}
where the prime denotes, the derivatives of potentials with
respects to their arguments.

\section{Massless case}

In the massless phase of gravity we have
$\langle{\phi^{a}}\rangle=0$, and thus, $
\langle{g^{\phi}_{\mu\nu}}\rangle=0$ and
$g_{\mu\nu}=\eta_{\mu\nu}$. In this case the curvature scalar $R$
vanishes with the constraint on the potentials
\begin{equation}
F(0)-V(0)=0 \label{vacuum}
\end{equation}
This constraint is important for determining the behavior of
potentials at their vanishing arguments. The vacuum solution of
potentials will be discussed in the next section.

\section{Massive case}

In the Minkowski background the metric tensor $g_{\mu\nu}$, the
arguments $y$, and $f$ are as follows
\begin{equation}
g_{\mu\nu}=\eta_{\mu\nu},
\end{equation}
\begin{equation}
y=y_{*}=m^{4}\eta_{\mu\nu}\eta^{\mu\nu}=4m^{4}, \label{y}
\end{equation}
\begin{equation}
f=f_{*}=\det(m^{4}\eta_{\mu\nu})/\det(\eta_{\mu\nu})=m^{16}. \label{f}
\end{equation}

Contracting the equation(\ref{metricmotion}), we get

\begin{equation}
yV'(y)+2\left[
F(f)+2fF'(f)-V(y)\right]=-R
\end{equation}
where $R$ is the curvature scalar. From this equation of motion
$y$ can be solved as the function of potentials $F(f)$, $V(y)$ and
their first derivatives as follows
\begin{equation}
y=\frac{2\left[V(y)
-F(f)-2fF'(f)\right]-R}{V'(y)} \label{trace}
\end{equation}
we denote such a solution as $y_{*}$, and in this case $f$,
becomes $f_{*}=m^{16}$, by considering equations (\ref{y},
\ref{f}), and $R=0$ in the Minkowski background. Rewriting
equation (\ref{trace}) as the function of arguments $y_{*}$ and
$f_{*}$ we get
\begin{equation}
y_{*}=\frac{2\left[V(y_{*})
-F(f_{*})-2m^{16}F'(f_{*})\right]}{V'(y_{*})}=4m^{4}
\end{equation}
If we set
\begin{equation}
F(f)+2fF'(f)=\Lambda
\end{equation}
Accordingly, the solutions for the two potentials are as follows:
\begin{equation}
F(f)=\Lambda+\lambda{f^{-1/2}} \label{F}
\end{equation}
\begin{equation}
V(y)=\Lambda+\frac{R}{2}+\lambda{y^{2}} \label{V}
\end{equation}
where $\Lambda$ is the cosmological constant, and $\lambda$ is a
constant with respect to $f$, and $y$. These are the prime results
of our work.

Einstein's equation of motion satisfies the general covariance
principle since it has tensorial form. So the physical quantities
should be coordinate independent, under the
\begin{equation}
x^{\mu}\rightarrow x^{\mu}-\varepsilon^{\mu}
\end{equation}
coordinate transformation. Where $\varepsilon^{\mu}$ is
infinitesimal. The scalar fluctuation can be gauged away by using
the diffeomorphisms
\begin{equation}
\delta\phi^{a}=\nabla_{\mu}\phi^{a}\varepsilon^{\mu}
\end{equation}
\begin{equation}
\delta h_{\mu\nu}=\nabla_{\mu}\varepsilon_{\nu}+\nabla_{\nu}\varepsilon_{\mu}
\end{equation}
after breaking the diffeomorphisms spontaneously by setting
\begin{equation}
\delta\phi^{a}=0.
\end{equation}

Considering the linearized Einstein equation (field equations for
$h_{\mu\nu}$) propagating in the Minkowski background, we have
\begin{equation}
g^{\phi}_{\mu\nu}\equiv m^{4}\eta_{\mu\nu},
\end{equation}
\begin{equation}
\det g^{\phi}_{\mu\nu}\equiv -m^{16}
\end{equation}
\begin{equation}
y=m^{4}\eta_{\mu\nu}g^{\mu\nu}=4m^{4}-m^{4}h+......=y_{*}-m^{4}h+.........
\end{equation}
\begin{equation}
f=-m^{8}/g=m^{16}-m^{16}h+.........=f_{*}-m^{16}h+.........
\end{equation}
Hence the linearized form of equation (\ref{metricmotion}) reads:
\begin{equation*}
(R_{\mu\nu}-\frac{1}{2}g_{\mu\nu}R)=
\end{equation*}
\begin{equation*}
\frac{1}{4}\eta_{\mu\nu}\left[V(y_{*})-F(f_{*})-8m^{16}F'(f_{*})-4m^{8}V''(y_{*})-4m^{32}F''(f_{*})\right]h
\end{equation*}
\begin{equation}
+\frac{1}{2}\left[F(f_{*})+2m^{16}F'(f_{*})-V(y_{*})\right]h_{\mu\nu}   \label{linear2}
\end{equation}
For decoupling of the scalar ghost state, the first term of the
right hand side of equation (\ref{linear2}) must vanish, i.e.
\begin{equation}
V(y_{*})-4m^{8}V''(y_{*})-F(f_{*})-8m^{16}F'(f_{*})-4m^{32}F''(f_{*})=0
\end{equation}
We can write equation (\ref{linear2}) in more general form

\begin{equation*}
(\partial_{\alpha}\partial_{\nu}h^{\alpha}_{\mu}+\partial_{\alpha}\partial_{\mu}h^{\alpha}_{\nu}-\partial_{\mu}\partial_{\nu}h
-\partial^{\alpha}\partial_{\alpha}h_{\mu\nu}-\eta_{\mu\nu}\partial_{\beta}\partial_{\alpha}h^{\beta\alpha}+\eta_{\mu\nu}\partial^{\alpha}\partial_{\alpha}h)
\end{equation*}
\begin{equation}
=m^{2}_{g}\left[\xi\eta_{\mu\nu}h-h_{\mu\nu}\right]
\label{general2}
\end{equation}
by introducing
\begin{equation}
\xi=\frac{1}{2}-\frac{2m^{8}[V''(y_{*})+m^{24}F''(f_{*})]+3m^{16}F'(f_{*})}{V(y_{*})-F(f_{*})-2m^{16}F'(f_{*})}
\end{equation}
\begin{equation}
m^{2}_{g}=V(y_{*})-F(f_{*})-2m^{16}F'(f_{*}) \label{mass}
\end{equation}
where $m_{g}$ is the mass of graviton. As can be seen from the
eqs. (\ref{F}), and (\ref{V}), one can relates the gravitation
mass to that of scalar field as
\begin{equation}
m^{2}_{g}=\lambda{m^{-8}} \label{mass1}
\end{equation}
The $\lambda$ parameter must be positive. After eliminating the
scalar ghost, and vector ghost states, the equation of motion
(\ref{general2}) becomes
\begin{equation}
h=0
\end{equation}
\begin{equation}
\partial^{\mu}h_{\mu\nu}=0
\end{equation}
\begin{equation}
\partial^{\alpha}\partial_{\alpha}h_{\mu\nu}-m^{2}_{g}h_{\mu\nu}=0 \label{graviton}
\end{equation}
The equation of motion (\ref{graviton}) describes the massive
graviton (spin-2) particle without ghost state in the linearized
approximation. In this situation the equations of motions are in
the form of \cite{zurab2007} and \cite{durmush2009}, but the
equation (\ref{mass}) for Pauli-Fierz combination of mass term is
determined as eq. (\ref{mass1}). This result is different what is
found in Demir and Pak \cite{durmush2009}, and \cite{zurab2007}.
This shift is proportional to the mass of scalar field, and the
parameter $\lambda$. Furthermore, for the linearized solution
$F(f)$ takes the form
\begin{equation}
F(f)=\Lambda+\lambda{m^{-8}}\left[1-\frac{1}{2}h\right] \label{F2}
\end{equation}
which is a function of the ghosty scalar $h=h^{\mu}_{\mu}$. Also,
to determine the nature of potentials, for example, the
non-trivial of the vacuum solution, with the form of $F(f)$, one
relates the two potentials to the cosmological constant,
$F(0)=V(0)\equiv\Lambda$. Hence, one can determine the vacuum
phase of the potentials. This relation between two potentials
enables one to kill the cosmological term with equation
(\ref{vacuum}). It is also the two potentials are equal in the
flat Minkowski space-time,
\begin{equation}
V(y_{*})-F(f_{*})=0. \label{F3}
\end{equation}
but different in the curved one.
\section{Conclusion}

We have analyzed massive graviton in the set up of Demir and Pak
\cite{durmush2009} with an explicit solution for the determinantal
potential. From our work, one can conclude that:

a) If the potential is as in equation (\ref{F}) the contribution
of determinantal potential to the Pauli-Fierz mass term becomes as
eq.(\ref{mass1}), this is very different from the works of
\cite{durmush2009}, and \cite{zurab2007}.

b) The structure of the determinantal potential enables us to
relate kinetic and determinantal potentials such that
$F(0)=V(0)=\Lambda$, the cosmological constant. However,
consistency of the theory for the massless case (equation
(\ref{vacuum})) guarantees that $\Lambda=0$. This may be used as
an argument for vanishing cosmological term in the set-up
considered.

\end{document}